\documentstyle[preprint,prd,aps,epsf]{revtex}
\begin{document}
\draft
\title{Nonequilibrium Photons as a Signature of \\
Quark-hadron Phase Transition}
\author{Da-Shin Lee\footnote{E-mail address: {\tt dslee@cc3.ndhu.edu.tw}}}
\address{Department of Physics, National Dong Hwa University, Hua-Lien, 
Taiwan, R.O.C.}
\author{Kin-Wang Ng\footnote{E-mail address: {\tt nkw@phys.sinica.edu.tw}}}
\address{Institute of Physics, Academia Sinica, Taipei, Taiwan, R.O.C.}
\date{December 1998}
\maketitle

\begin{abstract}
We study the nonequilibrium photon production in the quark-hadron
phase transition, using the Friedberg-Lee type solitons as a working 
model for quark-hadron physics. We propose that to search for nonequilibrium
photons in the direct photon measurements of heavy-ion collisions may be
a characteristic test of the transition from the quark-gluon to hadronic phases.
\end{abstract}
\vspace{2pc}
\vspace{2pc}

Quark-confinement is an unresolved problem in QCD physics. Nevertheless,
it is generally believed that quarks are deconfined at high energies. 
One of the primary goals of relativistic heavy-ion-collision
experiments such as BNL-AGS and CERN-SPS is to create a hot central region 
and a dense fragmentation region in which 
quark-gluon plasma (QGP) may be formed. 
One of the ways for testing the occurrence of QGP is to observe the spectrum
of the emitted photons. An advantage of measuring the direct photons 
is that photons do not suffer from strong final-state interactions as hadrons
do, so they can be used to monitor the initial stages of 
the colliding heavy ions~\cite{won}.

In relativistic heavy-ion collisions, a large number of $\pi^0$ and $\eta$
are produced by soft QCD processes. 
Their two-photon decays constitute the major source of photon emission.
Other photon sources arise from radiative decays of other mesons and baryon
resonances, as well as hadron scattering processes. 
Also, if QGP is formed, photons will be produced via quark-antiquark 
annihilations or quark-gluon Compton scatterings. 
It is an experimental challenge to separate out the photon yield into a part
arising from the decays of produced $\pi^0$ and $\eta$ and the 
"single photon" part from other sources. 
For instance, the part from decays of mesons can be subtracted by 
reconstructing the distribution of the two-photon invariant mass of all photon
pairs. If this can be done, 
then the excess single photon spectrum can be a probe of the properties
of the nuclear matter or QGP, or even a discriminator between the two phases.
However, recent studies~\cite{sri,sol,li} have shown 
that the two single photon spectra are similar, and
distinguishing between them might need further high-precision direct photon
measurements or other complementary measurements such as dilepton production
and $J/\psi$ suppression.

In this Letter, we give an attempt to utilize the nonequilibrium
nature of the photon emission associated with the quark-hadron phase transition
(QHPT) to test the formation of QGP. 
In essence, we calculate the particle production
from the release of latent heat of the phase transition, adopting the formalism
of nonequilibrium field theory that is well-suited to study the dynamics
of nonequilibrium processes~\cite{boy1,boy2,boy3}. 
To model the QHPT, we use the Friedberg-Lee (FL) 
phenomenological nontopological solition model~\cite{fri}, 
which is simple and adequate 
enough for our present considerations. In fact, FL model has been applied
to fit the hadronic properties~\cite{wil}, 
as well as to discuss the QHPT of the early Universe~\cite{sze,cot}. 
We believe that the particle production
considered here is generic to all Landau-Ginzberg type models of QHPT. 
Our main result will be the photon spectrum produced from the QHPT, 
and the experimental signature will be briefly discussed.

\begin{figure}
\leavevmode
\hbox{
\epsfxsize=6.5in
\epsffile{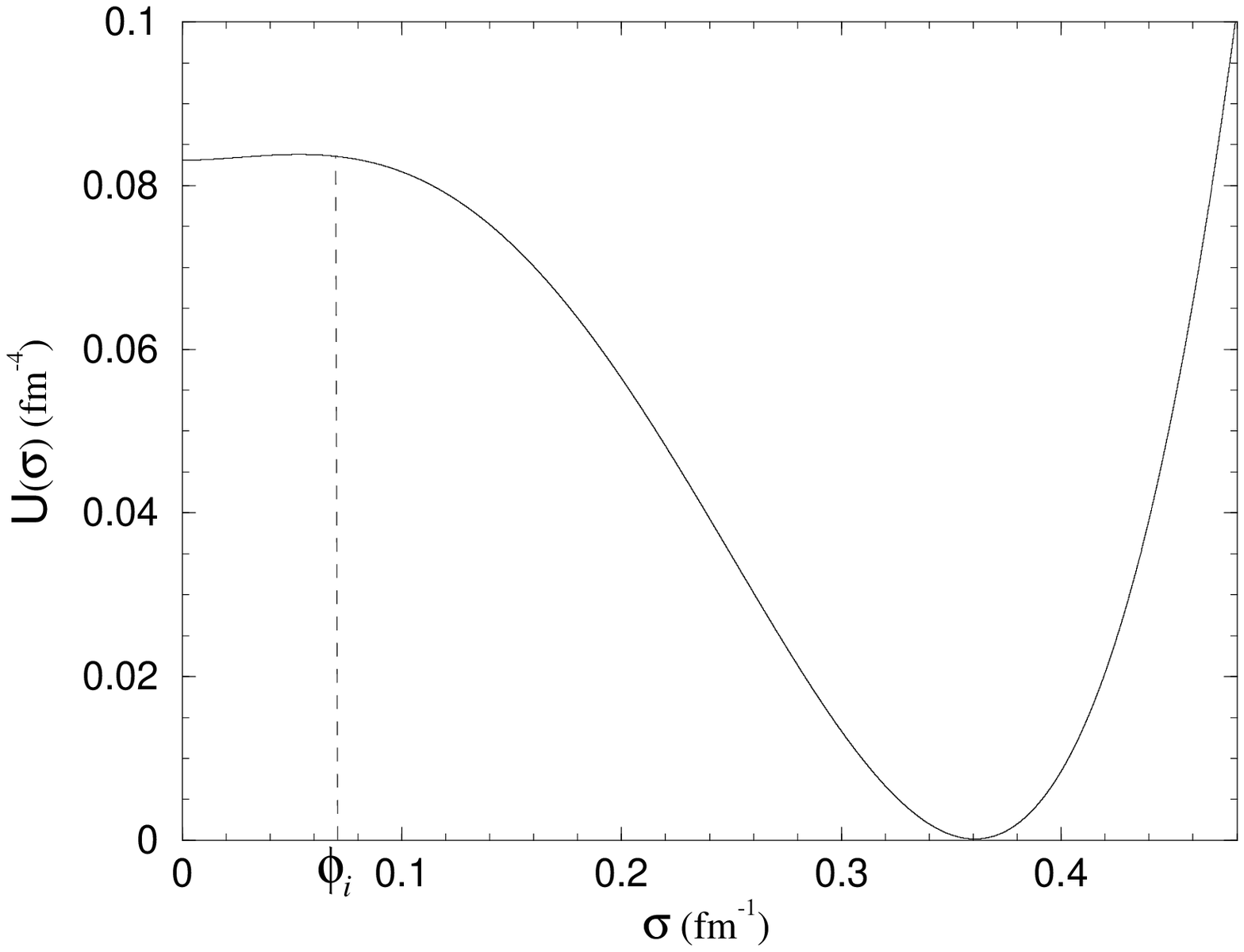}}
\caption{}
\label{fig1}
\end{figure}

We begin with the simplest FL model whose Lagrangian is given by~\cite{fri}
\begin{equation}
{\cal L}_\sigma = i\sum_{i=1}^{n_F}\bar\psi_i\gamma^\mu\partial_\mu\psi_i
   + {1\over2}\partial_\mu\sigma\partial^\mu\sigma - U(\sigma) -
   \sum_{i=1}^{n_F}f\sigma\bar\psi_i\psi_i, 
\end{equation}
where $\psi_i$ represents the quark field, $n_F=2$ is the number of 
light flavors, $\sigma$ is a real scalar field, $f$ is a coupling constant,
and the potential $U(\sigma)$ is given by 
\begin{equation}
U(\sigma)={a\over2!}\sigma^2-{b\over3!}\sigma^3+{c\over4!}\sigma^4+B,
\label{U}
\end{equation}
where $a$, $b$, and~$c$ are positive parameters, and $B$ is the bag constant.
Note that $U(\sigma)$ has a local minimum at $\sigma=0$, and an absolute 
minimum at $\sigma=\sigma_c$ which denotes the true vacuum with
$U(\sigma_c)=0$, separated by a potential barrier.
A typical potential is shown in Fig.~\ref{fig1}.
Due to the nonlinearity of $U(\sigma)$, the model bears nontopological soliton
solutions which are identified with hadrons. Essentially, inside a hadron is 
a perturbative vacuum with $\sigma\simeq 0$ decorated with localized free
quarks, whereas outside is the true vacuum. 

The simple picture of the production of nonequilibrium photons
from the QHPT is the following. Suppose that
a thermalized QGP is produced initially in a relativistic nucleus-nucleus
collision, which has a perturbative vacuum with $\sigma=0$. 
It then expands and cools, and undergoes hadronization in 
a phase transition. Here we assume that the transition is weakly
first-order as consistent with lattice QCD calculations~\cite{uka}.
As such, transition to a state with $\sigma\simeq\phi_i$ (see Fig.~\ref{fig1})
by quantum tunneling (or thermal activation) will occur through nucleation of 
hadronic bubbles in the QGP. Due to the plasma expansion,  
it is conceivable that the $\sigma$ field inside
a bubble might get trapped at $\sigma\simeq\phi_i$ as the interaction rate of 
$\sigma$ is small compared to expansion rate.
When the expansion slows down, $\sigma$ will start to oscillate and 
relax via production of the particles, 
to which the $\sigma$ particle is coupled, to the true vacuum.  
This nonequilibrium stage will take place for a time-scale of about 10 fm, 
as revealed by hydrodynamical simulations (for example, see Ref.~\cite{sol}
and references therein).
Typically, during this stage 
the dominant channel for the relaxation is the production
of strong interacting particles such as $\sigma$ quanta, gluons and other
mesons. These particles will be immediately trapped and thermalized 
in the nuclear matter. Perhaps the $\sigma$ quanta are short-lived  
to decay into photons before thermalization. If this happens, the
decay photons would be a source for nonequilibrium photons.  
However, our current interest is the direct 
photon production driven by the oscillations of the $\sigma$ field 
due to parametric amplification.
We will ignore the hydrodynamic expansion and simplify the dynamics by
adopting a "quenched" phase transition from an initial state in local
thermodynamic equilibrium at a temperature slightly above the critical 
temperature cooled instantaneously to zero temperature. This "quench"
approximation is far from a definitive description of the dynamics. 
Nevertheless, it captures the qualitative features and allows a simple 
but concrete calculation~\cite{boy2,boy3}. 

At tree level the $\sigma$ field is inert to electromagnetic interaction, 
but it can couple to photon through a quark loop. After integrating out
the quark field, we obtain an one-loop effective Lagrangian
including strong interacting final states,
\begin{eqnarray}
{\cal L}&=&{1\over2}\partial_\mu\sigma\partial^\mu\sigma - U(\sigma)
           -{1\over4} F_{\mu\nu}F^{\mu\nu} 
             -{1\over4} G_{\mu\nu}G^{\mu\nu} +{1\over2} m_V^2 B_\mu B^\mu
                  \nonumber \\
          && -\frac{g}{4\sigma_c}(\sigma-\sigma_c) F_{\mu\nu}F^{\mu\nu}
             -\frac{h}{4\sigma_c}(\sigma-\sigma_c) G_{\mu\nu}F^{\mu\nu},
\label{L0}
\end{eqnarray}
where $F_{\mu\nu}=\partial_\mu A_\nu - \partial_\nu A_\mu$ is the photon, 
and $G_{\mu\nu}=\partial_\mu B_\nu - \partial_\nu B_\mu$ is a 
quark-antiquark vector meson with mass $m_V$. 
It is straightforward to calculate the coupling constant 
$g \simeq 2.6 \times 10^{-3}$. But the coupling
constant $h$ would depend on the wave function 
at the origin of the vector meson, $\Psi_V(0)$.
Assuming $m_V\simeq 2\;{\rm GeV}$ and taking a conservative value for 
$|\Psi_V(0)|^2 \simeq 0.1\;{\rm fm}^{-3}$~\cite{wil}, 
we find $h\simeq 1.8 \times 10^{-2}$.
Note that effective strong interacting vertices such as 
$\sigma gg$ (where $g$ denotes a gluon) and $\sigma\pi\pi$ should also appear
at one-loop level.
However, we have omitted them in the effective Lagrangian~(\ref{L0})
because their effects to the particle production can be fully represented
by the $\sigma$ strong self-couplings. 
Since we are concerned with photon production only, we integrate out the 
vector meson and obtain
\begin{equation}
{\cal L}={1\over2}\partial_\mu\sigma\partial^\mu\sigma - U(\sigma)
                   -{1\over4} F_{\mu\nu}F^{\mu\nu} 
             -\frac{g}{4\sigma_c}(\sigma-\sigma_c) F_{\mu\nu}F^{\mu\nu}
              -\frac{h^2}{8m_V^2\sigma_c^2}
             \partial_\mu\sigma\partial_\alpha\sigma F^{\mu\nu}F^\alpha_{~\nu},
\label{L}
\end{equation} 
where we have dropped off higher derivative terms. 
At this stage, we should emphasize that the above obtained
 effective Lagrangian for the processes such as 
$\sigma' \rightarrow 2 \gamma $  as well as 
$ 2\sigma' \rightarrow 2 \gamma  $ ( here $\sigma'$ is the shifted field with
$ \sigma=\sigma'+ \sigma_c$)  is clearly understood in perturbation theory. 
It is not certain whether this effective Lagrangian can also describe 
the nonequilibrium situation. In other words, the effective vertices 
that account for the above mentioned  processes
may be modified in the strongly out-of-equilibrium situation.  
The fuller study of  off-equilibrium effective vertices  
by integrating out the quark fields and vector meson is a challenging  
task that lies beyond the scope of this paper, but certainly deserves 
to be taken up in the near future.  However, in this work, 
we will use this effective Lagrangian (\ref{L}) to compute 
the nonequilibrium photon production.
 
Following the nonequilibrium closed time path formalism\cite{boy1,boy2,boy3}, 
the nonequilibrium effective Lagrangian is given by
\begin{equation}
{\cal L}_{neq}={\cal L}\left[\sigma^+,A_\mu^+\right] -
               {\cal L}\left[\sigma^-,A_\mu^-\right],
\label{Lneq}
\end{equation}
where + (-) denotes the forward (backward) time branches. 
We then split $\sigma^{\pm}$ into a mean field and 
the quantum fluctuating fields:
\begin{equation}
\sigma^{\pm}(\vec x,t)=\phi(t) + \chi^{\pm}(\vec x,t),
\end{equation}
with the tadpole conditions,
\begin{equation}
\langle \chi^{\pm}(\vec x,t) \rangle = 0.
\label{tad}
\end{equation}
This tadpole conditions will be imposed to all orders in the corresponding 
expansion to obtain the nonequilibrium equations of motion 
from the first principles approach.

To derive the nonequilibrium evolution equations that consistently 
take into account quantum fluctuation effects 
from the strong $\sigma$ self-interaction, 
we adopt the following  
Hartree factorization for $\chi$  
 implemented for both $\pm$ components\cite{boy1,boy2,boy3}:
\begin{eqnarray}
&& \chi^4  \rightarrow 6 \langle \chi^2 \rangle \chi^2 + \rm{constant}, 
\nonumber \\
&& \chi^3 \rightarrow 3 \langle \chi^2 \rangle \chi. 
\end{eqnarray}
The expectation value will be determined self-consistently.
Before proceeding any further, it is worth noting that although  
the above Hartree factorization is uncontrolled in this effective   
field theory that involves a single scalar  field\cite{boy1,boy2,boy3}, 
our justification of using this approximation is based on the fact 
that it provides a non-perturbative framework that allows us to 
treat the strong $\sigma$ dynamics self-consistently. 

After doing this factorization, the Lagrangian then becomes
\begin{eqnarray}
{\cal L}\left[\phi(t)+\chi^+,A_\mu^+\right] &&-
               {\cal L}\left[\phi(t)+\chi^-,A_\mu^-\right] 
=\left\{ \frac{1}{2} (\partial\chi^+)^2- U (t) \chi^+ 
        -\frac{1}{2} M_{\chi}^{2}(t) \chi^{+2}
            -{1\over4}F^{+}_{\mu\nu}F^{+\mu\nu} \right. \nonumber \\
   &&      -\frac{g}{4\sigma_c}(\phi(t)-\sigma_c) F^+_{\mu\nu}F^{+\mu\nu}
    -\frac{g}{4\sigma_c} \chi^+  F^{+\mu\nu}F^+_{\mu\nu} 
      -\frac{h^2}{8m_V^2\sigma_c^2} \dot\phi^2(t)
          F^{+0i}F^{+0}_{~~~i}    \nonumber \\
&&  -\frac{h^2}{4m_V^2\sigma_c^2} 
\dot\phi(t) \partial_\alpha\chi^+
         F^{+0i}F^{+\alpha}_{~~~i}     \left.-\frac{h^2}{8m_V^2\sigma_c^2}   
        \partial_\mu\chi^+\partial_\alpha\chi^+  
        F^{+\mu\nu}F^{+\alpha}_{~~~\nu}  \right\} 
-\left\{ + \rightarrow -\right\}, \nonumber \\
\end{eqnarray} 
where 
\begin{eqnarray}
 U(t) &=& \ddot\phi(t)+ \left[a-{b\over2}\phi(t)+{c\over6}\phi^2(t)+\frac{c}{2}
                     \Sigma(t)\right] \phi(t)
          -{b\over2}\langle\chi^2\rangle(t),  \nonumber \\
 M_{\chi}^{2}(t) &=& a-b\phi(t)+{c\over2}\phi^2(t)+{c\over2}\Sigma(t), 
                      \nonumber \\
 \Sigma(t)&=& \langle\chi^2\rangle(t) -\langle\chi^2\rangle(0).
\end{eqnarray}
Here, we have performed a subtraction of $\langle\chi^2\rangle(t)$ at $t=0$ 
absorbing $\langle\chi^2\rangle(0)$ into the finite renormalization of $a$.
Besides, since we are 
interested in the processes of direct photon production driven by a time 
dependent $\phi(t)$ field, in which the photons do not appear
in the intermediate states, to avoid the gauge ambiguities, 
we can work on the Coulomb gauge, and concentrate only on physical transverse
gauge fields, $\vec A_T$\cite{boy3}.

With the above Hartree-factorized Lagrangian in the Coulomb gauge, 
we perform a perturbative expansion in the weak couplings $g$ and $h^2$. 
However,
the strong $\sigma$ dynamics is treated non-perturbatively\cite{boy3}. 
Following from the tadpole conditions~(\ref{tad}), 
we obtain  the following full one-loop equation of motion 
of $\phi(t)$ given by  
\begin{eqnarray}
&&\ddot\phi(t) + \left[a-{b\over2}\phi(t)+{c\over6}\phi^2(t)+\frac{c}{2}
                     \Sigma(t)\right] \phi(t)- 
\frac{g}{2\sigma_c}\langle|\partial_\mu\vec A_T|^2\rangle(t) 
- \nonumber \\
&&~~~~~~~~~~~~~~~~~~~~~~~~~~~~~\frac{h^2}{4m_V^2\sigma_c^2}
\left[\ddot\phi(t)+\dot\phi(t)\frac{d}{dt}\right]
\langle|\dot{\vec A_T}|^2\rangle(t)=0.
\label{eom}
\end{eqnarray}
The Heisenberg field equations can be read off from the quadratic part 
of the Lagrangian in the form
\begin{eqnarray}
&&\left[\frac{d^2}{dt^2}-\vec\nabla^2 +M_{\chi}^{2}(t)\right]\chi(t)=0, 
\nonumber \\
&&\left[1+\frac{g}{\sigma_c}(\phi(t)-\sigma_c)+\frac{h^2}{4m_V^2\sigma_c^2}
\dot\phi^2(t)\right]\ddot{\vec A_T}(t) 
+
 \left[\frac{g}{\sigma_c}\dot\phi(t) + \right.\nonumber \\
&&
\left. \frac{h^2}{2m_V^2\sigma_c^2}
\dot\phi(t)\ddot\phi(t)\right]\dot{\vec A_T}(t) - 
\left[1+\frac{g}{\sigma_c}(\phi(t)-\sigma_c)\right]
\vec\nabla^2\vec A_T(t) =0. 
\end{eqnarray}
Now we decompose the fields $\chi$ 
and $\vec A_T$ into their Fourier
mode functions $U_{\vec k}(t)$ and $V_{\lambda \vec k}(t)$ respectively,
\begin{eqnarray}
\chi(\vec x,t)&=&\int\frac{d^3k}{\sqrt{2(2\pi)^3\omega_{\chi\vec k}}}
\left[a_{\vec k} U_{\vec k}(t) e^{i\vec k\cdot \vec x} + {\rm h.c.}\right],
\nonumber \\
\vec A_T(\vec x,t)&=&\sum_{\lambda=1,2} \int
\frac{d^3k\;\vec \epsilon_{\lambda \vec k}}{\sqrt{2(2\pi)^3\omega_{A\vec k}}}
\left[b_{\lambda \vec k} V_{\lambda \vec k}(t) e^{i\vec k\cdot \vec x} +
{\rm h.c.}\right],
\end{eqnarray}
where $a_{\vec k}$ and $b_{\lambda \vec k}$ are destruction operators, and
$\vec \epsilon_{\lambda \vec k}$ are linear polarization unit vectors.
The frequencies $\omega_{\chi\vec k}$ and $\omega_{A\vec k}$ 
can be determined from the initial states and will be specified below.
Then the mode equations are
\begin{eqnarray}
&&\left[\frac{d^2}{dt^2}+k^2 +M_{\chi}^{2}(t)\right]U_k(t)=0, \nonumber \\
&&\left\{\left[1+\frac{g}{\sigma_c}(\phi(t)-\sigma_c)
+\frac{h^2}{4m_V^2\sigma_c^2}
\dot\phi^2(t)\right]\frac{d^2}{dt^2} \right. 
+\left[\frac{g}{\sigma_c}\dot\phi(t) + \right. \nonumber \\
&&
\left. \frac{h^2}{2m_V^2\sigma_c^2}
\dot\phi(t)\ddot\phi(t)\right] \frac{d}{dt} + 
k^2 \left[1+\left. \frac{g}{\sigma_c}(\phi(t)-\sigma_c)\right]\right\} 
V_{\lambda k}(t)=0, 
\label{meq}
\end{eqnarray}
with the vacuum expectation values given by
\begin{eqnarray}
\langle\chi^2\rangle(t)&=&
\int^\Lambda\frac{d^3k}{2(2\pi)^3\omega_{\chi k}}|U_k(t)|^2, \nonumber \\
\langle|\partial_\mu\vec A_T|^2\rangle(t)&=&
\sum_\lambda \int^\Lambda \frac{d^3k}{2(2\pi)^3\omega_{A k}}
\left[|\dot V_{\lambda k}(t)|^2-k^2|V_{\lambda k}(t)|^2\right],
\label{qf}
\end{eqnarray}
where we set the cutoff scale $\Lambda\simeq m_V$, and the expectation
value of the number operator for the asymptotic photons with momentum 
$\vec k$ is given by\cite{boy3}
\begin{eqnarray}
\langle {\bf N}_{k}(t)\rangle&=&
 {\frac{1}{2k}} \left[ \dot{\vec A_T}( {\vec k},t) \cdot 
\dot{\vec A_T}( -{\vec k},t) \right. \left.
+k^2  {\vec A_T}( {\vec k},t) \cdot 
{\vec A_T}( -{\vec k},t) \right]-1 \nonumber \\  
&=& {1\over 2k^2}\sum_\lambda
\left[|\dot V_{\lambda k}(t)|^2+k^2
| V_{\lambda k}(t)|^2\right] - 1.
\end{eqnarray}
This gives the spectral number density of the photons produced at time $t$,
$dN(t)/d^3k$.
To solve the evolution equations~(\ref{eom},\ref{meq}), 
we propose the following initial conditions for the mode functions 
at the time of "quench"\cite{boy3}:
\begin{eqnarray}
&&U_k(0)=1,\;\dot U_k(0)= -i\omega_{\chi k},
\;\omega_{\chi k}^2 = k^2+a+b\phi_i+{c\over2}\phi_i^2\;; \nonumber \\
&&V_{\lambda k}(0)=1,\;\dot V_{\lambda k}(0)= -i\omega_{A k},
\;\omega_{A k}=k,
\label{ic}
\end{eqnarray}
where the initial mode functions are chosen to be at zero temperature 
inside the hadronic bubbles  with the mean field $\phi(t)$  displaced 
initially away from the equilibrium position (i.e., $\phi (0) =\phi_i \ne 0$).
The above specified initial conditions are physically plausible and  
simple enough for us to investigate a quantitative description of the dynamics. 

Let us use the FL model parameters: $a=1.6\;{\rm fm^{-2}}$, 
$b=69\;{\rm fm^{-1}}$, $c=500$, and $f=9.57$. This implies that
$\sigma_c=0.36\;{\rm fm^{-1}}$. This set of parameters
has been used for fitting hadronic mass spectrum~\cite{wil}, 
and with this potential~(\ref{U}) (see Fig.~\ref{fig1}) the phase transition
is weakly first-order at finite temperature~\cite{sze}.
We choose the initial amplitude $\phi_i=0.07\;{\rm fm^{-1}}$ 
to solve Eq.~(\ref{eom}).

\begin{figure}
\leavevmode
\hbox{
\epsfxsize=6.5in
\epsffile{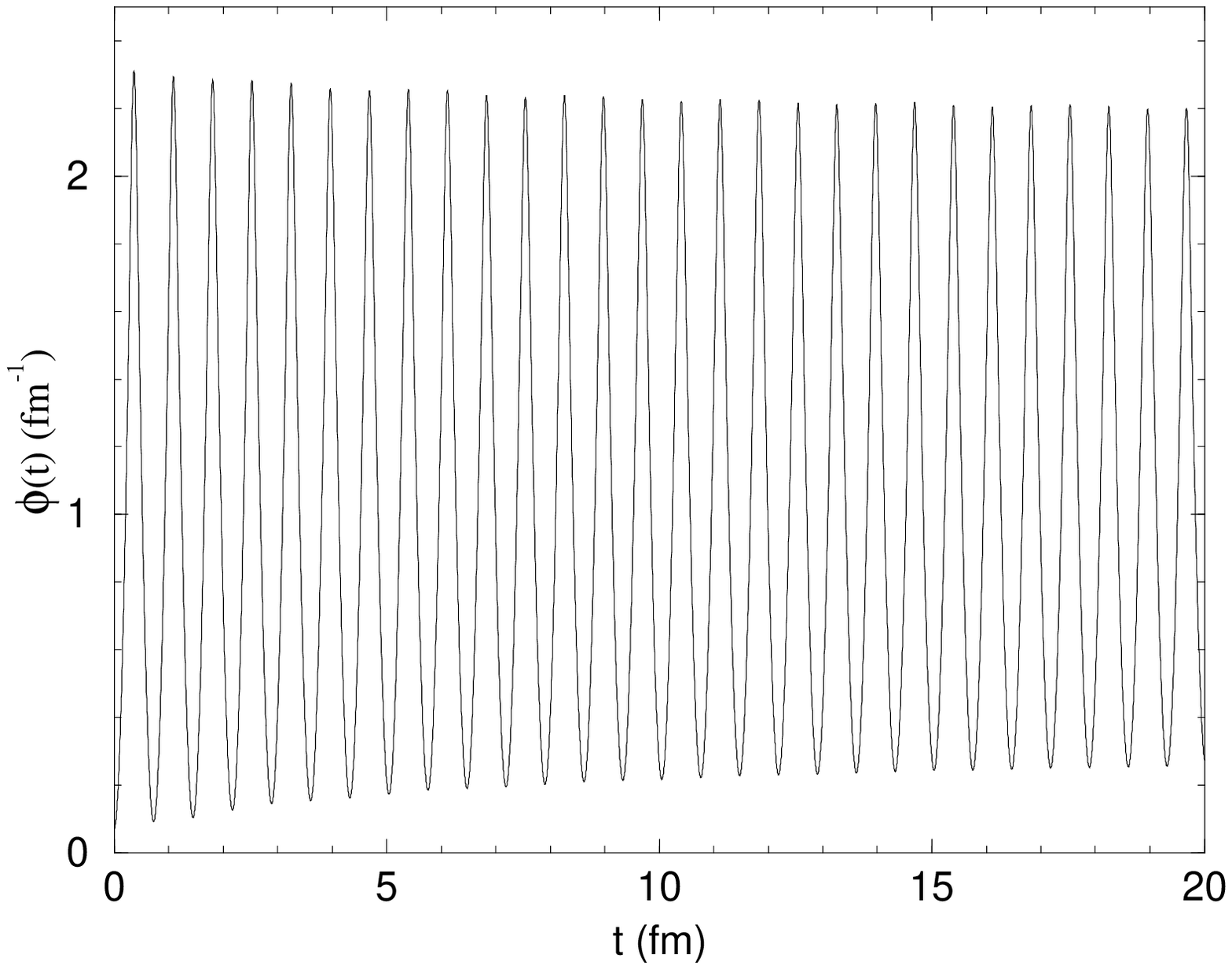}}
\caption{}
\label{fig2}
\end{figure}

\begin{figure}
\leavevmode
\hbox{
\epsfxsize=6.5in
\epsffile{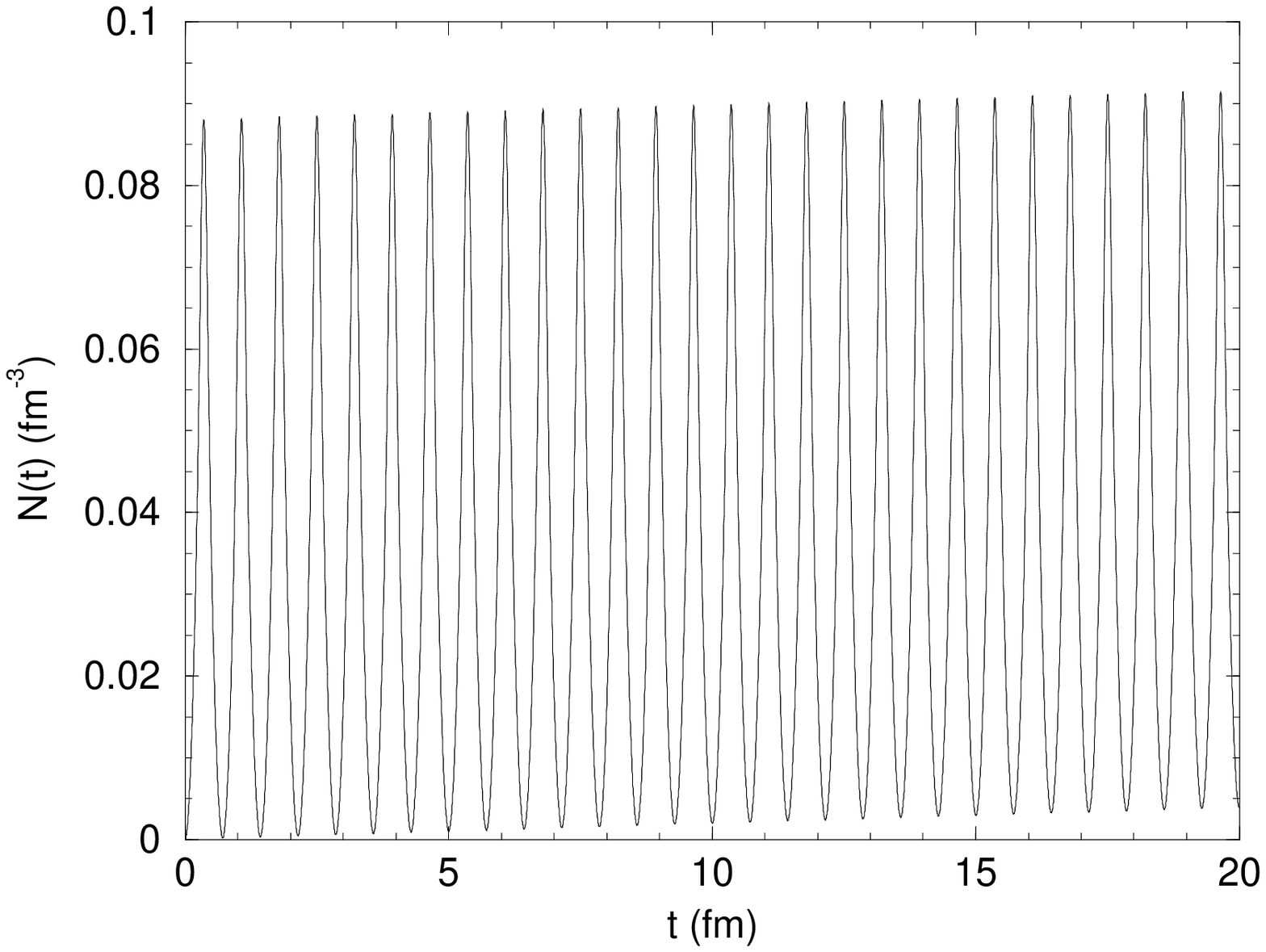}}
\caption{}
\label{fig3}
\end{figure}

In Fig.~\ref{fig2}, the time evolution of the mean field $\phi(t)$ is shown. 
It can be seen that $\phi$ is oscillating about a mean value of about 
$1.1\;{\rm fm^{-1}}$, which is significantly different from the classical
value $\sigma_c$. Also, $\phi(t)$ oscillates
with a frequency $\omega_\phi \simeq 8.8\;{\rm fm^{-1}}$.
This is due to  the quantum fluctuation effects coming from 
the $\sigma$ field and the gauge field $\vec A_T$  
to the motion of $\phi(t)$~(\ref{eom}). 
The back reactions of the quantum fields also 
account for the damping of $\phi(t)$ with time.
In Fig.~\ref{fig3}, we plot
the time dependence of the photon number density integrated over 
momentum~${\vec k}$, $N(t)$. 
As expected, $N(t)$ oscillates
with the same frequency as $\phi(t)$. Although photons are produced 
efficiently during the first half of an oscillation, almost all of them are
re-absorbed to the background during the second half of the oscillation. 
However, it is important to note that the time average of
$N(t)$ is indeed increasing with time, that is to say, 
photons are effectively produced from $\phi$ oscillations 
due to parametric amplification\cite{boy2}.

\begin{figure}
\leavevmode
\hbox{
\epsfxsize=6.5in
\epsffile{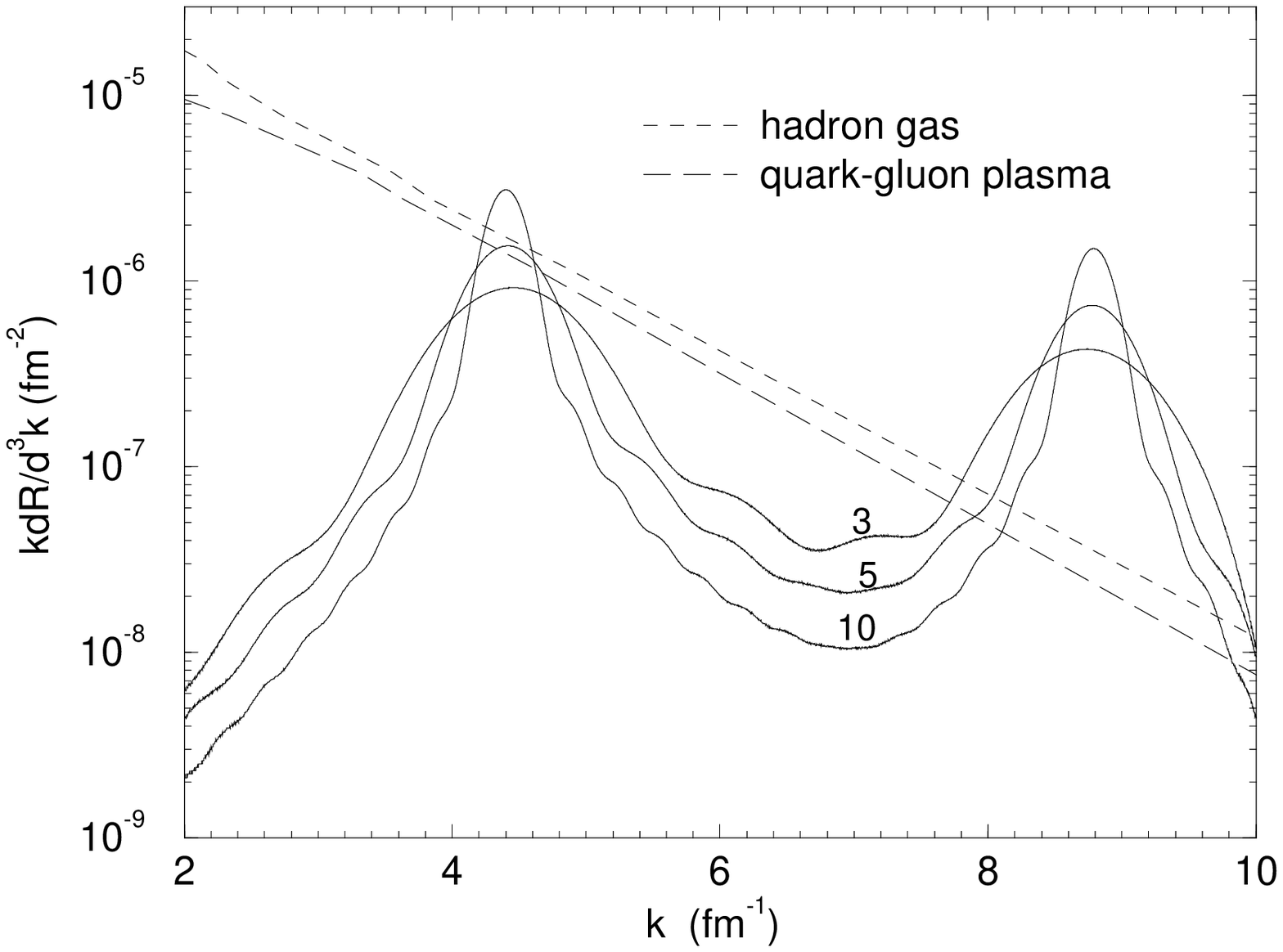}}
\caption{}
\label{fig4}
\end{figure}

It is useful to calculate the time-averaged invariant photon production rate,
$kdR/d^3k$, where 
\begin{equation}
dR={1\over T}\int_0^T \frac{dN(t)}{dt} dt,
\label{dR} 
\end{equation}
over a period from the initial time to time $T$.
The results are shown in Fig.~\ref{fig4} with $T=3, 5, 10\;{\rm fm}$.
Also shown are the 
thermal photon production rate from a quark-gluon plasma and a 
hadron gas taken from Ref.~\cite{kap}. 
It is apparent that the $\phi$ oscillations with frequency $\omega_\phi$  
produce non-thermal photons
whose spectrum peaks around two photon momenta, $k=\omega_\phi/2$ 
and $k=\omega_\phi$, which correspond to the unstable bands (or resonant bands) 
arising from the fact that  the mode equations of $\vec{A}_{T}$ (\ref{meq}) 
consistently  depend on the time dependent mean field $\phi(t)$ \cite{boy3}.
Clearly, the  photon production mechanism is that of  parametric amplification. 
From the mode equations (\ref{meq}), we can easily recognize  
that  the $4.4\;{\rm fm^{-1}}$ peak is resulted 
from  the coupling $\sigma F^2$ while 
the $8.8\;{\rm fm^{-1}}$ peak is from the interaction
$(\partial\sigma)^2 F^2$. Note that the $4.4\;{\rm fm^{-1}}$ peak has
a peak value of the production rate being comparable
to that of thermal photons, while interestingly the $8.8\;{\rm fm^{-1}}$ peak 
is almost two orders of magnitude larger than the thermal photons. Although
these results more or less depend on the choice of the FL 
model parameters, we do not find a change of two orders of magnitude.
So these high-energy non-thermal photons can be a distinct signature 
of QGP formation. 

In summary, we have computed the nonequilibrium photon production 
during the quark-hadron phase transition at high-temperature 
in the Friedberg-Lee model of quark-hadron physics. 
Under the "quench" approximation,
the invariant production rate for nonequilibrium photons driven by the 
oscillation of the $\sigma$ field due to parametric amplification is given, 
which is two orders of magnitude larger than that from a thermal quark-gluon 
plasma for photon energies around $2\;{\rm GeV}$. These high-energy non-thermal
photons can be a potential test of the formation of quark-gluon plasma in   
relativistic heavy-ion-collision experiments.
Of course, in order to compare the results with experimental data 
on direct photons, a more realistic dynamics of the phase transition should
be considered and the nonequilibrium photon production rate 
should be convolved with the expansion of the plasma. 
This work is in progress.   

We would like to thank D. Boyanovsky and S.-P. Li for their useful discussions. 
The work of D.S.L. (K.W.N.) was supported in part by the 
National Science Council, 
ROC under the Grant NSC88-2112-M-259-001 (NSC88-2112-M-001-042).

\newpage

\begin{center}
{\bf FIGURE CAPTIONS}
\end{center}
\bigskip
\noindent
Fig.~\ref{fig1}. Scalar potential $U(\sigma)$ with $a=1.6\;{\rm fm^{-2}}$, 
$b=69\;{\rm fm^{-1}}$, and $c=500$, and $\phi_i$ is the initial value.
\medskip

\noindent
Fig.~\ref{fig2}. Time evolution of the mean field $\phi(t)$.
\medskip

\noindent
Fig.~\ref{fig3}. Time evolution of produced total number density of 
nonequilibrium photons from quark-hadron phase transition.
\medskip

\noindent
Fig.~\ref{fig4}. Spectral production rate of nonequilibrium 
photons from quark-hadron phase transition, calculated from Eq.~(\ref{dR})
with $T=3, 5, 10\;{\rm fm}$, drawn with solid lines. Dashed lines are
the rates for hadron gas and quark-gluon plasma taken from Ref.~\cite{kap}.

\newpage

\newpage

\newpage

\newpage

\end{document}